\documentstyle[preprint,prc,aps,epsfig]{revtex}
\begin{document}
\draft
\title{Electron Quasielastic Scattering at High Energy from $^{56}$Fe, What Suppression?}
\author{K.S. Kim$^{1)}$\thanks{kyungsik@color.skku.ac.kr}
and L.E. Wright$^{2)}$}
\address{
1)BK21 Physics Research Division and Institute of Basic Science,
Sungkyunkwan University, Suwon, 440-746, Korea \\
2) Institute of Nuclear and Particle Physics, Ohio University,
Athens, OH 45701}

\maketitle
\begin{abstract}
Quasielastic electron scattering $(e,e')$ from $^{56}$Fe is
calculated at large electron energies (2-4 GeV) and large three
momentum transfer (0.5-1.5 GeV/c). We use a relativistic
mean-field single particle model for the bound and continuum
nucleon wavefunctions based on the $\sigma-\omega$ model and we
include the effects of electron Coulomb distortion in the
calculation. The calculations are compared to high energy data
from SLAC and more recent data from Jefferson Laboratory,
particularly for kinematics where the energy transfer is less
than 500 to 600 MeV and the quasielastic process is expected to
dominate the cross section. The effects of the predicted
weakening of the strong scalar and vector potentials of the
$\sigma-\omega$ model at high energy are investigated.  Possible
evidence for `longitudinal suppression' or modifications of
nucleon form factors in the medium is considered, but neither is
necessary to explain the quasielastic data for four momentum
transfers less than 1 (GeV/c)$^2$.

\end{abstract}
 \narrowtext

Medium and high energy electron scattering is one of the most
useful tools to study  nucleon properties inside nuclei,
especially in the quasielastic region using the inclusive
reaction $(e,e')$. There have been many experiments
\cite{mezi,dead,froi,hott,barr,alte,bates,sacl1} on medium and
heavy nuclei at incident electron energies less that 1 GeV, and a
number of theoretical attempts
\cite{jin1,bouc,trai,donn,hori,kim1,kim4} to fit the measured
cross section and to separate the longitudinal and transverse
structure functions. The Fermi gas model in the impulse
approximation provides a rough description of the inclusive
$(e,e')$ cross sections but fails in describing the details. This
is not too surprising since the Fermi gas model does not include
the nuclear spatial geometry correctly.   In some cases, there
appeared to be large suppression (about 30-40 \%) of the
longitudinal structure functions (as compared to theoretical
calculations of $(e,e')$ using non-relativistic wavefunctions and
current operators) which implied missing strength in the Coulomb
sum rule\cite{mezi}. There was also disagreement between the
transverse structure functions extracted from the experimental
data and the predictions of the Fermi gas model, but this was
expected since exchange currents, pion production, and other
processes induced by transverse photons were not included in the
model. Many attempts were made to explain this purported missing
strength of the longitudinal structure function by improving the
nuclear bound states, modifying the nucleon form factors in the
nuclear medium, including final state interactions, and
relativistic dynamics effects.

Two ingredients enter the comparison of experimental $(e,e')$
data from medium and heavy nuclei to theory.  One of these is the
inclusion of electron Coulomb distortion effects and the second
is the model used to calculate the nuclear transition current. In
the early 90's, Coulomb distortion for the reactions $(e,e')$ and
$(e,e'p)$ in the quasielastic region was treated exactly by the
Ohio University group\cite{bates,jin1,jin2,zhang,jin3} using
partial wave expansions of the electron wavefunctions. While such
Distorted Wave Born Approximation (DWBA) calculations permit the
comparison of different nuclear models against measured cross
sections and provide an invaluable check on various approximate
techniques of including Coulomb distortion effects, they are
numerically challenging and computation time increases rapidly
with higher incident electron energies.  Jourdan \cite{jourdan}
used the Ohio group's calculation of Coulomb distortion effects
to investigate the Coulomb sum rule. His conclusion was that the
sum rule appeared to be obeyed and thus there was no
`longitudinal suppression'.

In order to avoid the difficulties associated with DWBA analyzes
at higher electron energies and to look for a way to still define
structure functions, Kim and Wright \cite{kim1,kim4,kim2,kim3}
developed an approximate treatment of the Coulomb distortion. The
essence of the approximation is to include Coulomb distortion in
the four potential arising from the electron current by letting
the magnitude of the electron momentum include the effect of the
static Coulomb potential. This leads to an $r$-dependent momentum
transfer. This approximation allows the separation of the cross
section into a `longitudinal' term and a `transverse' term which
is not formally possible in a full DWBA calculation. For medium
and heavy nuclei at moderate incident electron energies, a good
treatment of Coulomb distortion effects is necessary in order to
extract the `longitudinal' and `transverse' structure functions.

 It should be noted that not all investigators
found a longitudinal suppression.  For example, the Ohio group
analyzed the Bates data \cite{bates} for $^{40}$Ca$(e,e')$ using
the relativistic $\sigma-\omega$  model mean field potential for
the bound and continuum nucleons along with the relativistic
current operator coupled with a good description of Coulomb
distortion. Their results were in good agreement with the data
with no evidence of longitudinal suppression.  This same model,
coupled with our approximate treatment of Coulomb distortion, was
compared to the Saclay quasielastic data on $^{208}$Pb taken with
both electrons and positrons.  The DWBA and approximate
calculations of Coulomb distortion by positrons and electrons
were not consistent with the data quoted by Saclay, so it was not
possible to extract a `longitudinal' structure function
\cite{kim4}. In addition, we investigated the
approximation\cite{trai} used by the Saclay group for Coulomb
corrections and found it not be a good approximation.

The absence or presence of `longitudinal suppression' has been
argued vigorously at various conferences--partially because of
different theoretical treatments, but also because of some
experimental discrepancies among various laboratories.  There is
now new $(e,e')$ data at much higher energies and momentum
transfer on a number of nuclei including $^{56}$Fe from JLAB
\cite{arring} which is similar kinematically to some older data
from SLAC \cite{slac}. To our knowledge, no one has attempted to
calculate the quasielastic contributions within a nuclear model
to these cross sections, probably because of the numerical
difficulties in calculating the $(e,e')$ process with a good
nuclear model at such high energies.  We have extended the
capabilities of our codes to handle these kinematics and in this
paper will compare our simple relativistic mean field model to
the available high energy data on $^{56}$Fe from both SLAC and
JLAB. In particular, we will look for kinematic regions at
relatively low energy transfer where quasielastic scattering is
expected to dominate the cross section.  Our results may be
useful in examining various scaling studies \cite{arring,slac} of
$(e,e')$ at large Q$^2$ and in separating the quasielastic process
from inelastic contributions.


In the plane wave Born approximation (PWBA), where electrons or
positrons are described as Dirac plane waves, the cross section
for the inclusive quasielastic $(e,e')$ processes can be written
as
\begin{equation}
\frac{d^2\sigma}{d\Omega d\omega}= \sigma_{M} \{
\frac{q^4_\mu}{q^4}  S_L(q,\omega) + [ \tan^2 \frac{\theta}{2} -
\frac{q^2_\mu}{2q^2} ]  S_T(q,\omega) \},  \label{pwsep}
\end{equation}
where $q_\mu ^2 = \omega^2-{\bf q}^2$ is the four-momentum
transfer, $\sigma _{M}$ is the Mott cross section and $S_L$ and
$S_T$ are the longitudinal and transverse structure functions
which depend only on the three-momentum transfer $q$ and the
energy transfer $\omega$. Explicitly, the structure functions for
a given bound state with angular momentum $j_{b}$ are given by
\begin{eqnarray}
S_{L}(q,{\omega})&=&\sum_{{\mu}_{b}s_{p}}{\frac {{\rho}_{p}}
{2(2j_{b}+1)}} \int {\mid}N_{0}{\mid}^{2}d{\Omega}_{p} \\
S_{T}(q,{\omega})&=&\sum_{{\mu}_{b}s_{p}}{\frac {{\rho}_{p}}
{2(2j_{b}+1)}} \int
({\mid}N_{x}{\mid}^{2}+{\mid}N_{y}{\mid}^{2})d{\Omega}_{p}
\end{eqnarray}
with the outgoing nucleon density of states ${\rho}_{p}={\frac
{pE_{p}} {(2\pi)^{2}}}$. The ${\hat {\bf z}}$-axis is taken to be
along the momentum transfer ${\bf q}$ and ${\mu}_{b}$ and $s_{p}$
are the z-components of the angular momentum of the bound and
continuum state nucleons. The Fourier transform of the nuclear
current $J^{\mu}({\bf r})$ is simply,

\begin{equation}
N^{\mu}=\int J^{\mu}({\bf r})e^{i{\bf q}{\cdot}{\bf r}}d^{3}r ,
\end{equation}
where $J^{\mu}({\bf r})$ denotes the nucleon transition current.
The continuity equation has been used to eliminate the
$z$-component ($N_{z}$) via the equation $N_{z}=-{\frac {\omega}
{q}}N_{0}$ which is valid if current is conserved.  Note that if
we do not have nuclear current conservation (i.e., a different
Hamiltonian in the initial and final state), we need to calculate
$N_{z}$ directly.

 The {\it ad} {\it hoc}
expressions for the longitudinal and transverse structure
functions with inclusion of the electron Coulomb distortion (see
Ref. \cite{kim4} for details) are similar to above, but the
Fourier operators are modified by Coulomb distortion.  We include
the Coulomb distortion effects in our results, but unlike the
medium energy cases, Coulomb effects on the cross section at
these high electron energies are quite small.

The nucleon transition current in the relativistic single particle
model is given by

\begin{equation}
J^{\mu}({\bf r})=e{\bar{\psi}}_{p}({\bf r}){\hat J}^{\mu}
{\psi}_{b}({\bf r}) \;,
\end{equation}
where ${\hat J}^{\mu}$ is a free nucleon current operator, and
$\psi_{p}$ and $\psi_{b}$ are the wave functions of the knocked
out nucleon and the bound state, respectively. For a free
nucleon, the operator consists of the Dirac contribution and the
contribution of an anomalous magnetic moment $\mu_{T}$ given by
${\hat J}^{\mu}=F_{1}(q_{\mu}^2){\gamma}^{\mu}+
F_{2}(q_{\mu}^2){\frac {i{\mu}_{T}}
{2M}}{\sigma}^{\mu\nu}q_{\nu}$. The form factors $F_{1}$ and
$F_{2}$  are related to the electric and magnetic Sachs form
factors $G_{E}$ and $G_{M}$ by $G_{E}=F_{1}+{\frac
{{\mu}_{T}q_{\mu}^{2}}{4M^{2}}}F_{2}$ and
$G_{M}=F_{1}+{\mu}_{T}F_{2}$ which are assumed to take the
following standard form \cite{hole}:

\begin{equation}
G_{E}={\frac {1} {(1- {\frac {q^{2}_{\mu}}{\Lambda^2})^{2}}}}
={\frac {G_{M}} {({\mu}_{T}+1)}} \;,
\end{equation}
where the standard value for $\Lambda^2$ is 0.71 (GeV/c)$^2$.
Several investigators \cite{cheon1,thom} have suggested that
nuclear medium effects may affect the value of $\Lambda^2$ and
the value of $\mu_T$.

Four $(e,e')$ data sets on $^{56}$Fe taken at SLAC and JLAB
correspond to significant cross sections for energy transfers
less than 500-600 MeV.  Three of these data sets were taken at
SLAC \cite{slac} with the following initial electron energy and
scattering angle: E$_i$=2.02 GeV, $\theta=15^o$,
E$_i$=2.02 GeV,$\theta=20^o$, and E$_i$=3.595 GeV, $\theta=16^o$
while the fourth was taken at JLAB \cite{arring} with: E$_i$=3.595
GeV,$\theta=16^o$. In Figs.~1-4 we show the experimental data as
compared to three theoretical results, with the longitudinal and
transverse contributions to the cross section being shown for the
results labeled ``Lorentz''.

Our standard calculation at lower energies is to use a current
conserving model where the bound and continuum nucleons move in
the scalar S(r) and vector V(r) potentials generated by the
TIMORA code \cite{horo}. This result is labeled ``Const. S \& V''
in the four figures. However, the Ohio State group\cite{clark}
found in their global fits to proton-nucleus scattering from a
range of nuclei that the strengths of the scalar and vector
potentials decreased as the proton energy increased. We chose to
investigate this effect, also considered in an approximate way in
Ref. \cite{KHF}, by using a parametrization of S and V strength
as a function of proton kinetic energy consistent with the results
that Cooper {\it et al.}\cite{clark} found.   In particular, we
calculated $(e,e')$ with S(r) and V(r) for protons and neutrons
scaled by the functions, $f_S= 0.97-0.66 x + 0.28 x^2$ and $f_V=
0.97-0.91x+ 0.30 x^2$ respectively where $x$ is the outgoing
nucleon kinetic energy divided by the nucleon mass ($x=T_p/M$).
For 500 MeV nucleons these factors are 0.70 and 0.59.  The
results using these weakened potentials for the outgoing nucleons
are labeled, ``$T_p$-dep. S \& V''. As noted earlier, changing
the potentials for the bound and continuum nucleons results in
current non-conservation.  To get an estimate of the size of this
effect, rather than using the relation $N_z=-\omega/q N_o$ from
current conservation we evaluated $N_z$ directly.  This result is
labeled ``Lorentz'' since it is calculated in the Lorentz gauge.

The peak of the quasielastic peak in Fig. 1 occurs at an energy
transfer of approximately $150$ MeV and hence contains very
little pion production or other inelastic processes.  Our
relativistic-mean field calculation with the energy dependent
scalar and vector potentials fit the data very well. Furthermore,
since the longitudinal contribution is a significant fraction of
the total, there is no evidence for any kind of longitudinal
suppression.   One very interesting consequence of using the
energy dependent scalar and vector potentials is the much more
rapid fall off of the quasielastic peak at higher energy transfer
as compared to the energy independent potentials.  Since the fit
at the peak is much better with the energy dependent potentials,
this result suggests that most of the cross section above the
quasielastic peak is due to inelastic processes.

In Figs.~2-4, the kinematics lead to the peak of the quasielastic
peak being well above pion production threshold and hence, it is
not surprising that the theoretical model falls 10-20\% below the
data on the low energy side of the quasielastic peak.  The most
troubling case for our model is the 2.02 GeV data from SLAC at
$20^o$ shown in Fig.~2 where our curve falls below the data on the
low energy side of the quasielastic peak for energy transfers
between 150 and 200 MeV. However, due to the electron kinematics,
most of the cross section is expected to be transverse. Again, the
energy dependent potentials predict a rapid fall off of the
quasielastic peak on the high energy side. Scaling models of these
data sets should take this result into account.

Various authors \cite{cheon1,thom} have suggested that the nucleon
form factors, $G_E$ and $G_M$ are modified in the nuclear
medium.  For example, in Ref. \cite{cheon1}, the $\Lambda^2$
factor in Eq. 6 is predicted to be 0.41 (Gev/c)$^2$ for $G_E$ and
0.58 (GeV/c)$^2$ for $G_M$ for the proton while the proton
anomalous magnetic moment is predicted to increase from 1.71 to
1.91 while the neutron anomalous magnetic moment changes from
-1.92 to -2.21. Since the longitudinal and transverse cross
sections are roughly proportional to the squares of these form
factors we can estimate the effect at Q$^2$=0.25 (GeV/c)$^2$
($G_E$ drops by 11 \%, while $G_M$ increases by 3\%) and at  1.0
(GeV/c)$^2$ ($G_E$ drops by 51\% and $G_M$ decreases by 16\%).
Such decreases are not needed in our fit in Fig.~1 and they would
imply much greater inelastic contributions to the results shown
in Figs.~2-4.

A recent experiment\cite{jones} from JLAB has shown that the ratio
of $G_E/G_M$ for the proton begins to decrease for Q$^{2}$ values
over about 0.50 (GeV/c)$^2$.  Assuming that $G_E$ is the form
factor that is decreasing, this would show up in the longitudinal
contribution to the cross section.  To see this in the data sets
shown in Figs.~2-4 would require a good theoretical modeling of
pion production which, at this low energy transfer, is the primary
inelastic process.

In conclusion, we find no evidence of any kind of `longitudinal
suppression' in quasielastic scattering from $^{56}$Fe at high
momentum transfer.  This agrees with our previous
analyses\cite{bates} of quasielastic scattering from $^{40}$Ca at
much lower energies and momentum transfer. That is, the
relativistic mean field of the $\sigma-\omega$ model with energy
dependent scalar and vector potentials provides an excellent
description of quasielastic scattering from medium and heavy
nuclei.  While it is certainly possible to use inadequate nuclear
models coupled with approximate current operators to deduce the
need for additional effects, Occam's razor would suggest that the
relativistic mean field description is superior.  In addition,
our results suggest that there is no evidence of modification of
the nucleon form factors in the nuclear medium.

\section*{Acknowledgments}
This work was partially supported by the Korea Research Foundation
Grant KRF-2001-015-DP0103 and the US Department of Energy under
Contract DE-FG02-93ER40756 with Ohio University.


\begin{figure}[p]
\caption{Theoretical quasielastic scattering cross sections for
$^{56}$Fe with incident electron energy of $E_i=2.02$ GeV and
electron scattering angle $\theta=15^o$as a function on energy
transfer compared to experimental data from SLAC.  The cross
sections labeled $\sigma_T$ and $\sigma_L$ add up to the model
cross section labeled `Lorentz'. See text for details of the
models.} \label{s202-15}
\end{figure}

\begin{figure}[p]
\caption{Same as in Fig. 1, except $\theta=20^o$} \label{s202-20}
\end{figure}

\begin{figure}[p]
\caption{Same as Fig. 1, except the electron energy $E_i=3.593$ 
GeV and the scattering angle $\theta=16^o$.} \label{s3595-16}
\end{figure}

\begin{figure}[p]
\caption{Same as Fig. 1, except that the electron energy
$E_i=4.045$ GeV, the scattering angle $\theta=15^o$, and the
experimental data is from JLAB. } \label{j4045-15}
\end{figure}

\begin{references}
\bibitem{mezi}Z. E. Meziani {\it et} {\it al}., Nucl. Phys. {\bf A446},
113 (1985); Phys. Rev. Lett. {\bf 52}, 2130 (1984); {\bf 54},
1233 (1985).
\bibitem{dead}M. Deady {\it et} {\it al}., Phys. Rev. C {\bf 33}, 1897
(1986); {\bf 28}, 631 (198); C. C. Blatchley, J. J. LeRose, O. E.
Pruet, P. D. Zimmerman, C. F. Williamson, and M. Deady, Phys.
Rev. C {\bf 34}, 1243 (1986).
\bibitem{froi}B. Frois, Nucl. Phys. {\bf A434}, 57 (1985).
\bibitem{hott}A. Hotta, P. J. Ryan, H. Ogino, B. Parker, G. A. Peterson,
and R. P. Singhal, Phys. Rev. C {\bf 30}, 87 (1984).
\bibitem{barr}P. Barreau {\it et} {\it al}., Nucl. Phys. {\bf A402}, 515
(1983); {\bf 358}, 287 (1981).
\bibitem{alte}R. Altemus, A. Cafolla, D. Day, J. S. McCarthy,
R. R. Whitney, and J. E. Wise, Phys. Rev. Lett. {\bf 44}, 965
(1980).
\bibitem{bates} C. F. Williamson {\it et} {\it al}., Phys. Rev. C {\bf56},
3152 (1997): T. C. Yates {\it et} {\it al}., Phys. Lett. B {\bf
312}, 382 (1993).
\bibitem{sacl1} A. Zghiche, {\it et} {\it al}., Nucl. Phys. {\bf A573},
513 (1994); P. Gu\`{e}ye, et al., Phys. Rev. C {\bf 60}, 044308
(1999).
\bibitem{jin1}Yanhe Jin, D. S. Onley, and L. E. Wright, Phys. Rev. C
{\bf 45}, 1333 (1992).
\bibitem{bouc}P. M. Boucher and J. W. Van Orden, Phys. Rev. C {\bf 43},
582 (1991).
\bibitem{trai}M. Traini, S. Turck-Chi\'eze, and A. Zghiche, Phys. Rev.
C {\bf 38}, 2799 (1988); Phys. Lett. B {\bf 213}, 1 (9188).
\bibitem{donn}T. W. Donnelly, J. W. Van Orden, T. de Forest, Jr.,
and W. C. Hermans, Phys. Lett. B {\bf 76}, 393 (1978).
\bibitem{hori}Y. Horikawa, F. Lenz, and N. C. Mukhopadhyay, Phys. Rev. C
{\bf 22}, 1680 (1980).
\bibitem{kim1}K. S. Kim, L. E. Wright, Yanhe Jin, and D. W. Kosik,
Phys. Rev. C {\bf 54}, 2515 (1996).
\bibitem{kim4}K. S. Kim, L. E. Wright, and D. A. Resler, Phys. Rec. C
{\bf 64}, 044607 (2001).
\bibitem{jin2}Yanhe Jin, D. S. Onley, and L. E. Wright, Phys. Rev. C
{\bf 45}, 1311 (1992).
\bibitem{zhang}  Yanhe Jin, J.K. Zhang, D.S. Onley and L.E. Wright,
Phys. Rev. C {\bf 47}, 2024 (1993).
\bibitem{jin3} Yanhe Jin, D.S. Onley and L.E. Wright, Phys. Rev. C
{\bf 50}, 168 (1994).
\bibitem{jourdan} J. Jourdan, Nucl. Phys. A{\bf 603}, 117 (1996).
\bibitem{arring} J. Arrington et al., Phys. Rev. Lett. {\bf 82}, 2056 (1999);
J. Arrington et al., Phys. Rev. C {\bf 64},
014602 (2001); J. Arrington, Ph.D. Thesis, Cal. Tech, 1998. URL:
http://www.krl.caltech.edu/~johna/thesis
\bibitem{slac} D.B. Day et al., Phys. Rev. Lett. {\bf 59}, 427
(1987).
\bibitem{kim2}K. S. Kim and L. E. Wright, Phys. Rev. C {\bf 56},
302 (1997).
\bibitem{kim3}K. S. Kim and L. E. Wright, Phys. Rev. C {\bf 60},
067604 (1999).
\bibitem{hole}G. H\"ohler, E. Pietarinen, I. Sabba-Stevanescu,
F. Borkowski, G. G. Simon, V. H. Walter, and R. D. Wendling,
Nucl. Phys. {\bf B114}, 505 (1976).
\bibitem{cheon1}Il-Tong Cheon and Moon Taeg Jeong, J. Phys. Soc.
Japan, {\bf 61}, 2726 (1992); Moon Taeg Jeong and Il-Tong Cheon,
Phys. Rev. D {\bf 43}, 3725 (1991).
\bibitem{thom}K. Saito, K. Tsushima, and A. W. Thomas, Phys. Lett.
B {\bf 465}, 27 (1999).
\bibitem{horo}C. J. Horowitz and B. D. Serot, Nucl. Phys.
{\bf A368}, 503 (1981).
\bibitem{clark} E.D. Cooper, S. Hama, B. Clark, and R.L. Mercer,
Phys, Rev. C{\bf 47}, 297 (1993).
\bibitem{KHF} Hungchong Kim, C.J. Horowitz, and M.R. Frank, Phys.
Rev. C{\bf 51}, 792 (1995).
\bibitem{jones} M.K. Jones {\it et al.}, Phys. Rev. Lett. {\bf
84}, 1398 (2000).
\end{references}
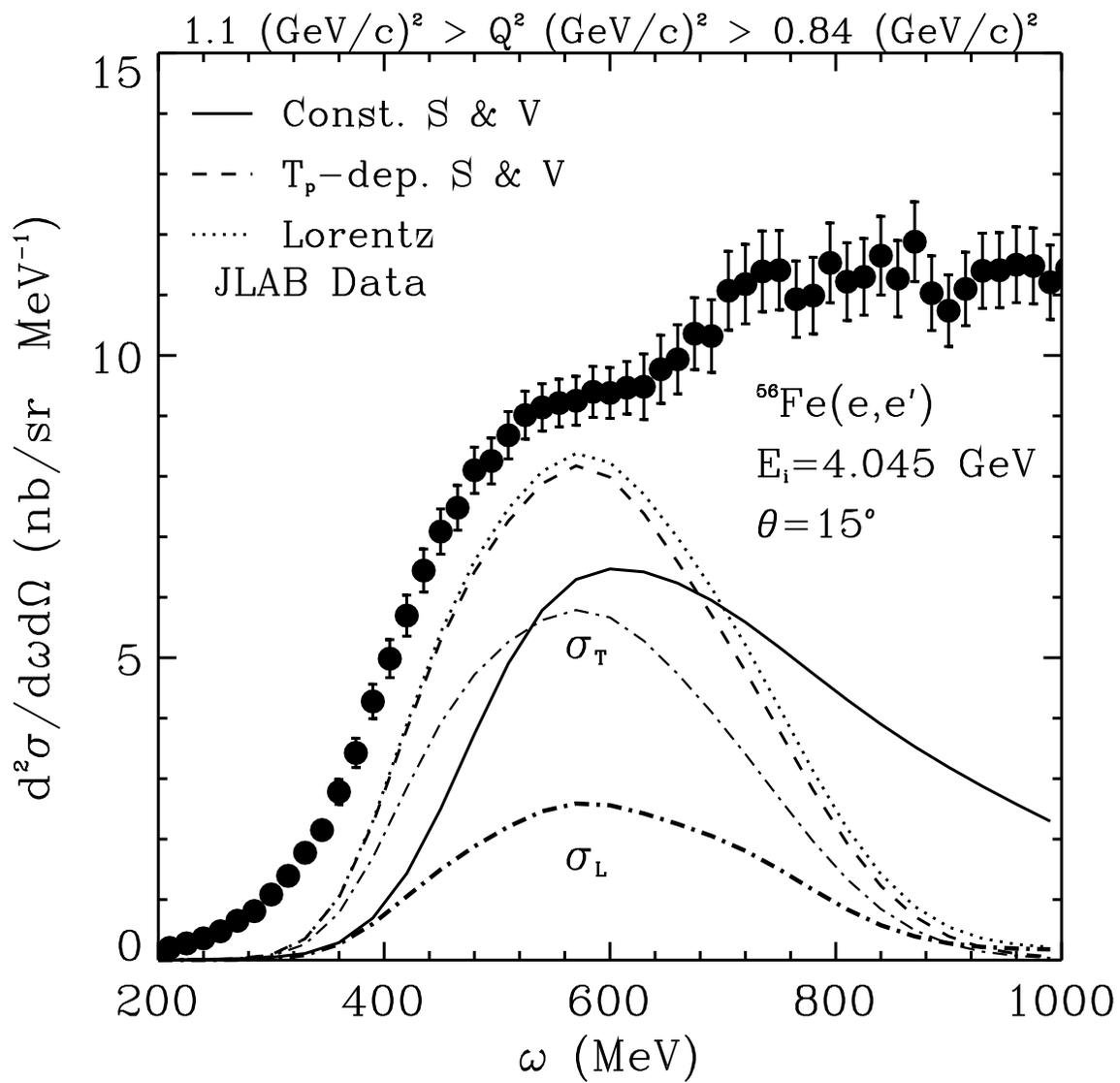
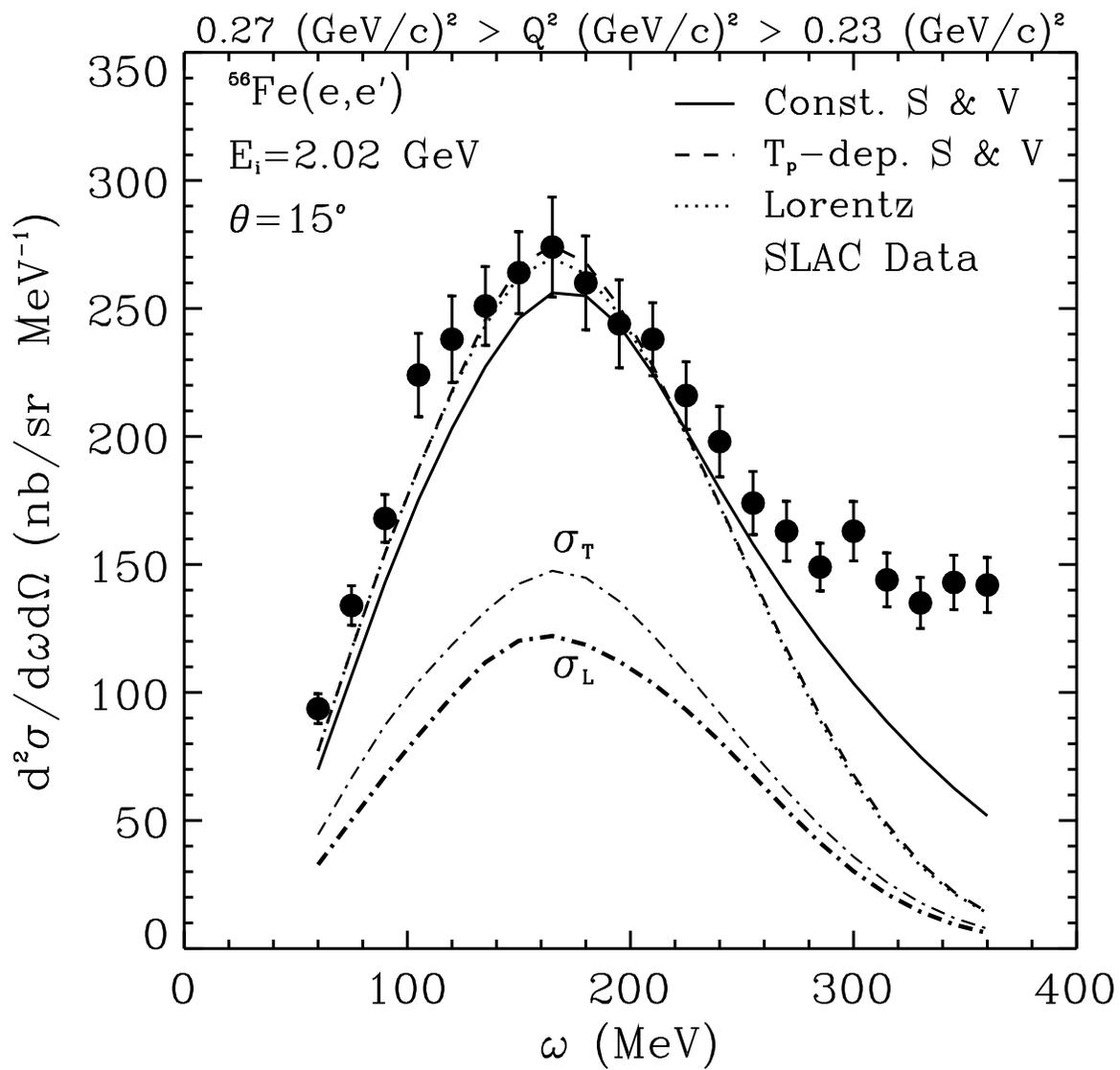
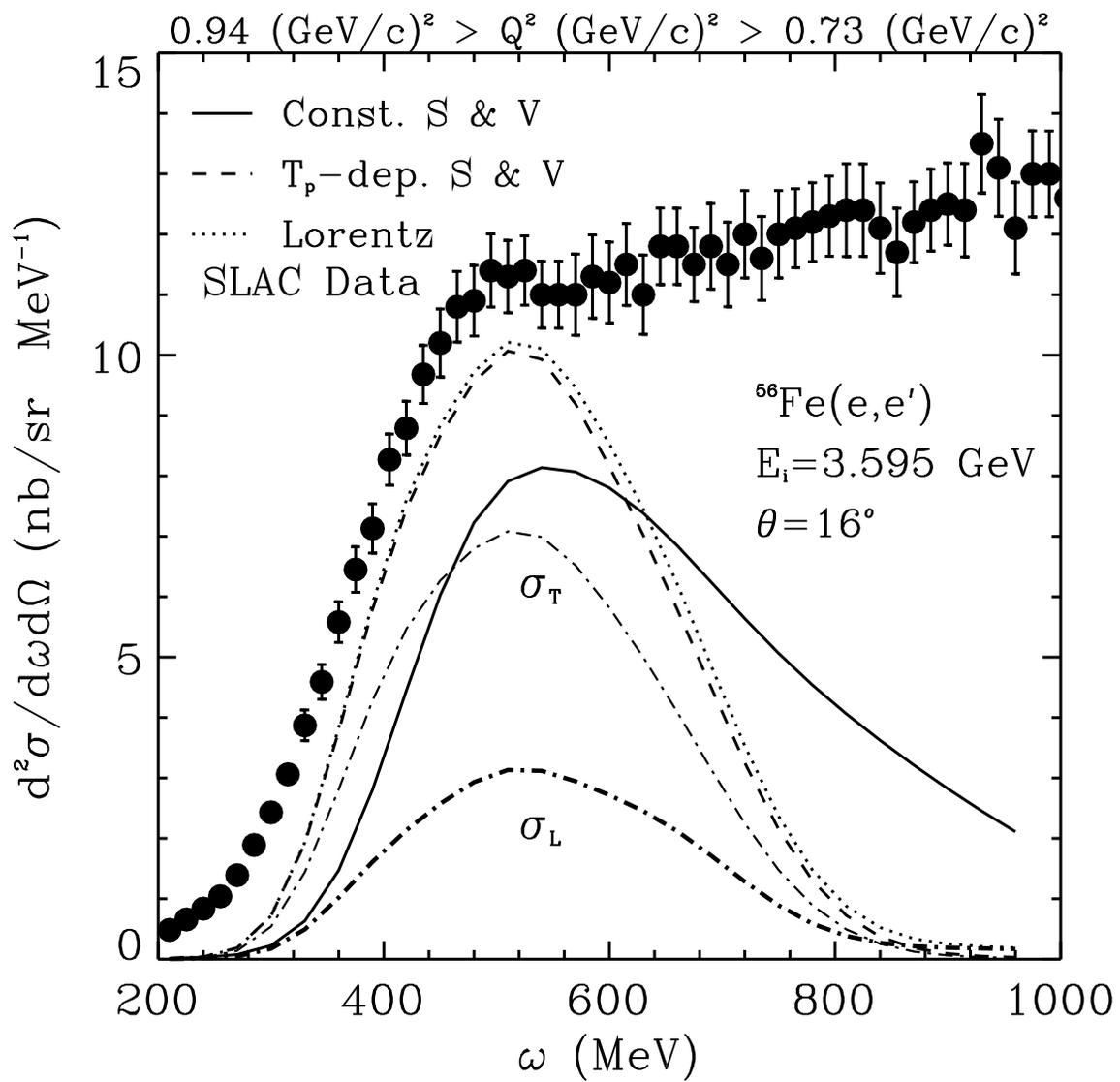
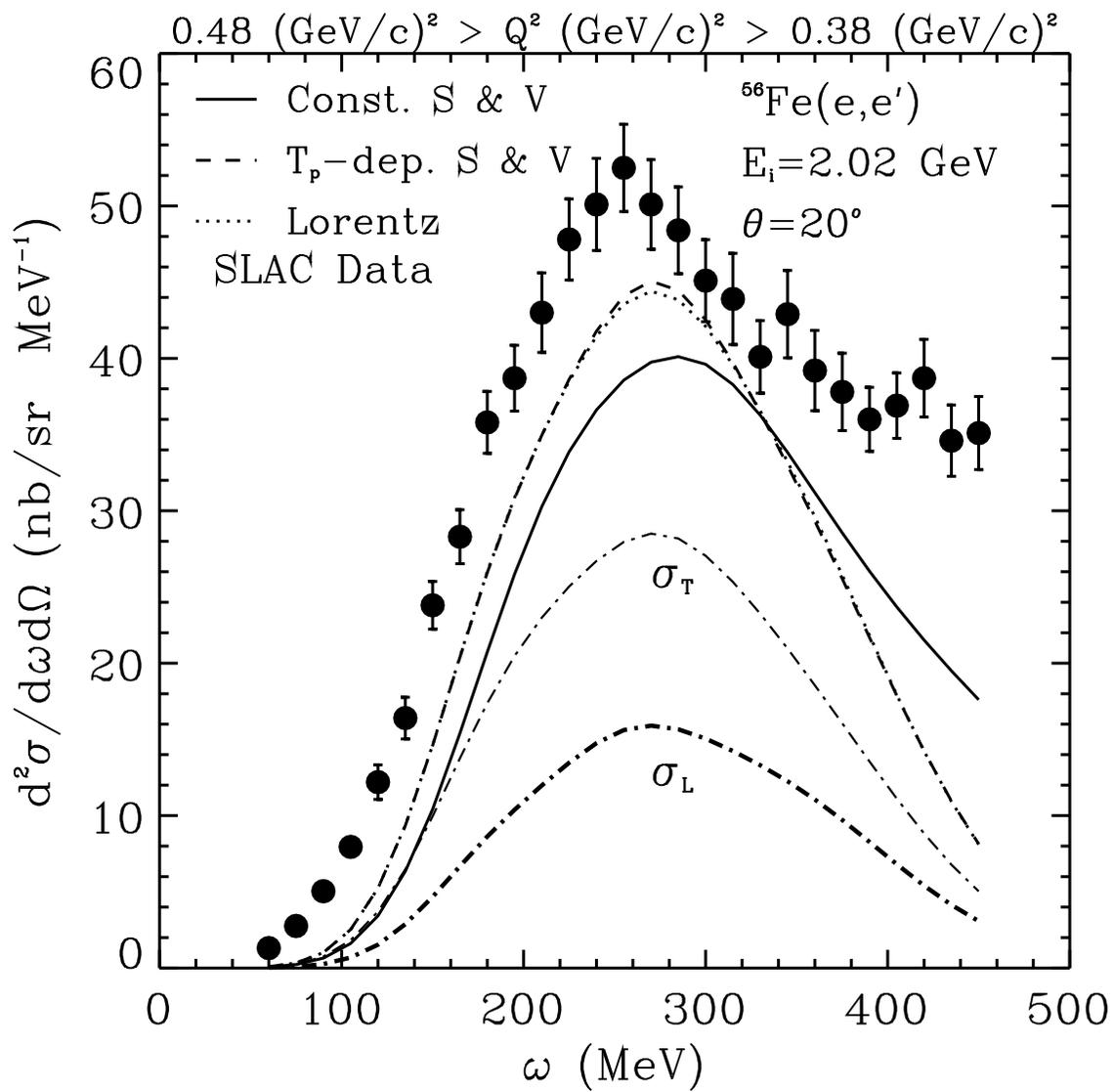
\end{document}